\documentclass[%
 reprint,
superscriptaddress,
 amsmath,amssymb,
 aps,
prd
]{revtex4-1}

\usepackage{graphicx}% Include figure files
\usepackage{dcolumn}% Align table columns on decimal point
\usepackage{bm}% bold math
\usepackage{graphicx}
\usepackage{wasysym}
\usepackage{amsfonts}
\usepackage{subfigure}
\usepackage{hyperref}
\usepackage{color}        
\usepackage{ulem}
\newcommand{\sfig}[2]{
\includegraphics[width=#2]{#1}
        }
\newcommand{\Sfig}[2]{
    \begin{figure}[ht]
    \sfig{#1.pdf}{0.95\columnwidth}
    \caption{{\small #2}}
    \label{fig:#1}
    \end{figure}
}
\newcommand{\Wfig}[2]{
    \begin{figure*}[ht]
    \sfig{#1.pdf}{0.95\textwidth}
    \caption{{\small #2}}
    \label{fig:#1}
    \end{figure*}
}
\newcommand{\chk}{$C_{h \kappa}$}
\newcommand{\chkh}{$C_{h \kappa}^{1h}$} 
\newcommand{\chkhh}{$C_{h \kappa}^{2h}$} 
\newcommand{\chh}{$C_{hh}$}
\newcommand{\ckk}{$C_{\kappa\kappa}$}

\begin{document}

\preprint{APS/123-QED}

\title{Detectability of Weak Lensing Modifications under Galileon Theories}% Force line breaks with \\

\author{Youngsoo Park}
% \email{youngsoo@uchicago.edu}
\affiliation{%
Department of Physics, University of Chicago, Chicago, Illinois 60637, U.S.A.
}%
\affiliation{%
Kavli Institute for Cosmological Physics, University of Chicago, Illinois 60637, U.S.A.
}%
% \altaffiliation[Also at ]{Physics Department, XYZ University.}%Lines break  automatically or can be forced with \\
\author{Mark Wyman}%
 %\email{mark.wyman@nyu.edu}

\date{\today}% It is always \today, today,

\begin{abstract}
Theories of modified gravity attempt to reconcile physics at the largest and the smallest scales by explaining the accelerated expansion of our universe without introducing the cosmological constant. One class of such theories, known as Galileon theories, predict lensing potentials of spherically symmetric bodies, such as dark matter halos, to receive a feature-like modification at the 5\% level. With the advent of next-generation photometric surveys, such modifications can serve as novel probes of modified gravity. Assuming an LSST-like fiducial dataset, we produce halo-shear power spectra for LCDM and Galileon scenarios, and perform a Fisher analysis including cosmological, nuisance, and Galileon parameters to study the detectability of the aforementioned modifications. With the LCDM scenario as our null hypothesis, we conclude that it is possible to detect the Galileon modifications at up to 4-$\sigma$ if present, or strongly exclude the model in a non-detection, with a tomography of four redshift bins and four mass bins, an LSST-like set of survey parameters, and Planck priors on cosmological parameters.

%\begin{description}
%use the optional argument of the \verb+\item+ command to give the category of each %item. 
%\end{description}
\end{abstract}

\pacs{Valid PACS appear here}% PACS, the Physics and Astronomy
\maketitle

%\tableofcontents

\section{Introduction}
\label{sec:intro}

The expansion history of our universe, especially its late-time acceleration as observed in \cite{Riess:1998cb,Perlmutter:1998np}, is explained within General Relativity (GR) as the effect of the cosmological constant. However, this approach creates a problem of vast differences in the predicted magnitude of the vacuum energy density between GR and Quantum Field Theory  \cite{RevModPhys.61.1}, known as the cosmological constant problem. This discrepancy, in turn, motivates infrared modifications of GR in an attempt to explain cosmic acceleration without introducing the cosmological constant.

The theories of massive gravity are some of the very few consistent ways of altering gravity on large length scales. These theories also motivate a new general class of scalar field theories, the Galileons \cite{Nicolis:2008in}. Unlike Galileons that one might think of from a usual field theory perspective, the Galileons that arise from theories involving a mass for gravitons \cite{deRham:2010ik,deRham:2010tw} have a new ``disformal" coupling to matter in the form of
\begin{equation}
\left[h_{\mu\nu} + \alpha \pi \eta_{\mu\nu}+\frac{\beta}{\Lambda_3^3}\partial_\mu\pi\partial_\nu\pi\right]T^{\mu\nu},
\end{equation}
where $\eta_{\mu\nu} = \text{diag}(-1,1,1,1)$, $h_{\mu\nu} = g_{\mu\nu}-\eta_{\mu\nu}$ for the metric tensor $g_{\mu\nu}$, $\alpha$, $\beta$ are dimensionless coefficients of order unity, $\Lambda_3^3 \equiv M_{Pl}m_g^2$ with $M_{Pl}$ and $m_g$ respectively being the Planck and graviton masses, $\pi$ the Galileon scalar field, and $T$ the stress-energy tensor. This coupling is different from the usual coupling of scalar fields to matter because of the third term. This term has many potential consequences, with one of them being a modification of gravitational lensing by spherically symmetric bodies. One of the authors found in \cite{MarkPRL} that such effects are exhibited as a 5\% level modification to the lensing potential of a spherically symmetric object. 

With the advent of state-of-the-art photometric surveys such as the Dark Energy Survey (DES) and the Large Synoptic Survey Telescope (LSST), it will be possible to use precise measurements of gravitational lensing to test modified theories of gravity. We investigate a realistic detection scenario for this modification within such future surveys by studying galaxy-galaxy lensing within LSST-like data to forecast how precisely deviations from GR can be detected. With a Fisher analysis of cosmological, nuisance, and Galileon parameters, we predict that future surveys could observe the new effect at up to 4-$\sigma$ if it is present, or put strong upper bounds on its presence.

The structure of this paper is as follows. In Section \ref{sec:theory}, a brief summary of the theoretical background is presented. In Sections \ref{sec:calc} and \ref{sec:results}, the component-by-component modeling of the stacked weak lensing signal, its modification under the Galileon model, and the statistical analyses on the detectability of the modification is described. Section \ref{sec:discussion} summarizes the results and discusses prospects for the near future.

\section{Theory}
\label{sec:theory}

In line with \cite{MarkPRL,Nicolis:2008in,deRham:2010ik,deRham:2010tw}, the Lagrangian for the scalar part of the Galileon theory is given by
\begin{eqnarray}
\mathcal{L}_\pi & = & \frac{3\eta}{2}(\partial \pi)^2+\frac{\mu}{\Lambda_3^3}(\partial \pi)^2\square \pi +\frac{\nu}{\Lambda_3}([\Pi]^2(\partial \pi)^2 - \nonumber \\
& & 2[\Pi]\partial_\mu \Pi_\nu^\mu\partial^\nu \pi - [\Pi^2](\partial \pi)^2 +2\partial_\mu \Pi_\nu^\mu \Pi_\lambda^\nu \partial ^\lambda \pi ) \nonumber \\
& & +(\alpha\pi\eta_{\mu\nu}+\frac{\beta}{\Lambda_3^3}\partial_\mu\pi\partial_\nu\pi)T^{\mu\nu}.
\end{eqnarray}
Here, $(\partial \pi)^2 \equiv \partial_\mu \pi \partial^\mu \pi$ and $\Pi_\nu^\mu \equiv \partial^\mu \partial_\nu \pi$. Three additional dimensionless parameters, $\eta$, $\mu$, and $\nu$, are introduced as well. Within the gravitational lensing context, a number of simplifying assumptions are made on this Lagrangian. We ignore the impact of the cosmological background value of the field $\pi$, assuming late times ($z<1$). We ignore time derivatives of $\pi$, assuming lenses that are not rapidly evolving. We ignore the impact of the  $\partial_\mu \pi \partial_ \nu \pi T^{\mu \nu}$ coupling on the solution for $\pi$, as it is a small perturbation.  Finally, we ignore deviations from spherical symmetry, following standard assumptions of lensing analysis. However, whether the average of random, non-spherically symmetric $\pi$ profiles reach spherical symmetry would be an important future work, since they are not simple linear or power--law functions. 

\Sfig{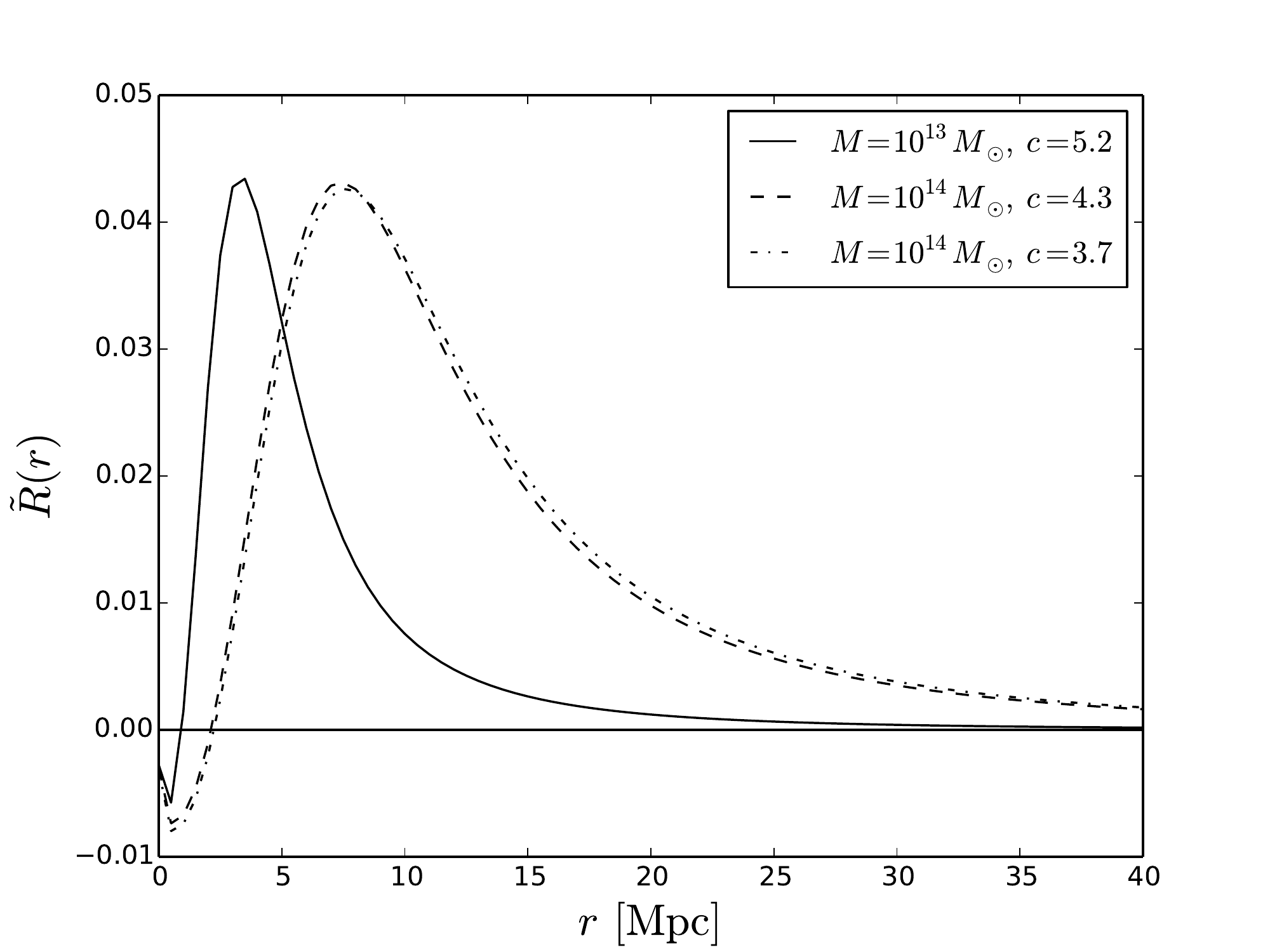}{Modification function $\tilde{R}(r)$, representing the fractional change in tangential shear generated by Galileon modification, for NFW halos of different masses and concentrations. Standard Galileon parametrization and scaling parameter $A_\pi=1$ are used.}

With these assumptions, the equations of motion for the scalar field $\pi$ yield the following analytic solution:
\begin{equation}
\label{eq:pi}
3\eta\left(\frac{\pi'}{r}\right) + \frac{4\mu M_P}{\Lambda_3^3}\left(\frac{\pi'}{r}\right)^2 + \frac{8\nu M_P^2}{\Lambda_3^6}\left(\frac{\pi'}{r}\right)^3 = \frac{\alpha G M(r)}{r^3}.
\end{equation}
There are five free parameters $\alpha$, $\beta$, $\eta$, $\mu$, $\nu$, and a reasonable starting point for the values of these parameters, motivated by massive gravity scenarios, is $\{\alpha, \beta, \eta, \mu, \nu\} = \{1, 1, 3, 6, 8\}$. The resulting Galileon modification to lensing potential is given by
\begin{eqnarray}
\label{eq:beta}
\Delta\Phi = \frac{\beta}{\Lambda_3^3}(\partial	 _r \pi)^2.
\end{eqnarray}

Let us denote the Galileon lensing potential as $\Phi_\text{Gal} \equiv \Phi_\text{GR}+\Delta\Phi$, with $\Phi_\text{GR}$ being the lensing potential in GR, and discuss how we translate the obtained modification of the lensing potential to a modification of an observable, namely the tangential shear. We  begin by assuming a Navarro-Frenk-White (NFW) profile \cite{NFW}, and plugging it into Eq. \ref{eq:pi}. This produces a solution for $\pi '$, which then yields a lengthy but closed-form solution for the modification of the lensing potential, and therefore a solution for $\Phi_\text{Gal}$. Then, for both GR and Galileon lensing potentials, we obtain the deflection potential $\Psi$ as
\begin{equation}
\Psi(\boldsymbol{\theta}) = \frac{D_\text{LS}}{D_\text{L}D_\text{S}}\frac{2}{c^2}\int \Phi (D_\text{L}\boldsymbol{\theta},z)dz,
\end{equation}
where $D_\text{LS}$, $D_\text{L}$, $D_\text{S}$ are the angular diameter distances between lens and the source, between the observer and the lens, and between the observer and the source, respectively. Here, $\boldsymbol{\theta}$ represents the projected position of the source with respect to the lens, such that the vector $D_\text{L}\boldsymbol{\theta}$ points from the lens to the source. From $\Psi$, the two shear components $\gamma_1$ and $\gamma_2$ are derived as
\begin{eqnarray}
\gamma_1(\boldsymbol{\theta}) & = & \frac{1}{2}\left(\frac{\partial^2\Psi(\boldsymbol{\theta})}{\partial \theta_1^2}-\frac{\partial^2\Psi(\boldsymbol{\theta})}{\partial \theta_2^2}\right), \\
\gamma_2(\boldsymbol{\theta}) & = & \frac{\partial^2\Psi(\boldsymbol{\theta})}{\partial \theta_1 \partial \theta_2},
\end{eqnarray}
where $\boldsymbol{\theta} = \left(\theta_1,\theta_2\right) =  \left(\theta \cos \phi, \theta \sin \phi \right)$. Finally, tangential shear $\gamma_t$ is given by
\begin{equation}
\label{eq:gamma12}
\gamma_t (\boldsymbol{\theta}) = -\gamma_1(\boldsymbol{\theta})\cos 2\phi -\gamma_2(\boldsymbol{\theta})\sin 2\phi.
\end{equation}
Section II of \cite{Jeong:2009wi} provides an excellent visualization for the derivation of $\gamma_t$ from $\gamma_1$ and $\gamma_2$.

Now, we need to compare the two $\gamma_t$'s, calculated for GR and Galileon potentials. We first note that both potentials are spherically symmetric, which allows for two useful simplifications. First, as $\gamma_t(\boldsymbol{\theta})$ does not depend on $\phi$, it is a function of solely the separation $\theta$, so we let $\gamma_t(\boldsymbol{\theta}) = \gamma_t(\theta)$. Second, we thus may freely choose a value of $\phi$ for calculation of $\gamma_t(\theta)$, so we make a simplifying choice of $\phi=0$ and obtain 
\begin{equation}
\gamma_t(\theta) = -\gamma_1(\theta,\phi=0),
\end{equation}
removing the second term in Eq. \ref{eq:gamma12}. With these simplifications, we define the fractional modification function $R(\theta)$ as 
\begin{equation}
\label{eq:modfunc}
R(\theta) = \frac{\gamma_{t,\text{Gal}}(\theta)}{\gamma_{t,\text{GR}}(\theta)} -1
= \frac{\gamma_{1,\text{Gal}}(\theta,\phi=0)}{\gamma_{1,\text{GR}}(\theta,\phi=0)} -1.
\end{equation}

Let us consider the parameter dependences of $R(\theta)$. It obviously depends on the five Galileon parameters, and also depends on the NFW parameters $M$ and $c$, the halo mass and concentration. Ideally, we would choose a Galileon parameter that acts as a well-behaved scaling parameter and include it in a Fisher analysis to study the detectability and degeneracies of this modification. In Eq. \ref{eq:pi}, we see $\alpha$ as a candidate for such a parameter, but in practice $\alpha$ does not behave as a trivial scaling parameter due to the cubic equation on the LHS. Also, in Eq. \ref{eq:beta}, we note that $\beta$ may serve as such a parameter. While this is in fact true, we note that varying $\beta$ implies effectively varying the mass of the graviton, and therefore choose to refrain from using $\beta$. Therefore, we resort to introducing an ad hoc, linear scaling parameter, $A_\pi$, of the modification, for inclusion in the Fisher analysis, by defining the scaled modification function $\tilde{R}(\theta)$ as
\begin{equation}
\tilde{R}(\theta) = A_\pi \left( \frac{\gamma_{t,\text{Gal}}(\theta)}{\gamma_{t,\text{GR}}(\theta)} -1 \right).
\end{equation}
This is a technique similar to the multiplicative bias parameter commonly used in weak lensing calibrations. This way, we recover GR at $A_\pi=0$, and obtain Galileon results at $A_\pi=1$, with a linear transition in between the two cases. Thus, a Fisher analysis including $A_\pi$ allows us to gauge how detectable a modification that follows the template of the Galileon $R(\theta)$, as illustrated in Fig. \ref{fig:plots/modfuncs}, will be.

\section{Power Spectra and Their Galileon Modifications}
\label{sec:calc}
The dataset we will be assuming is a set of observations of the tangential shear of background galaxies due to foreground galaxies (so-called galaxy-galaxy lensing). The core observable then is the cross-correlation between galaxy over-density (the positions of the foreground galaxies) and background shears. This is a 2-point function that can be modeled for any cosmology and -- with the aid of the modifications described in \S \ref{sec:theory} -- computed for any set of Galileon parameters. The noise on this measurement will of course be relevant for projections, and this too depends on a set of 2-point functions.

We begin then by generating a suite of fiducial power spectra under a standard Friedmann-Robertson-Walker (FRW) cosmology with cosmological constant and cold dark matter (LCDM) assumptions. Then, the Galileon modifications depicted in Fig. \ref{fig:plots/modfuncs} are propagated to changes in these spectra. Armed with these power spectra as a function of the scaling parameter $A_\pi$, we perform a Fisher analysis to study the detectability of the Galileon modifications, using the LCDM power spectra as the null hypothesis. This also means that we use the LCDM covariance of power spectra for the analysis.

We assume a stacked galaxy-galaxy lensing analysis. Using the halo model \cite{Peacock:2000qk, Seljak:2000gq} and a global mass profile for dark matter haloes, such as the aforementioned NFW profile, one may place lensing objects in mass bins and superimpose their lensing signals on top of each other to vastly improve observational statistics. Such measurements are well represented by both real space (mean tangential shear) and Fourier space (halo-shear power spectrum) observables. The latter provides a simpler approach to covariances and Fisher matrix calculations, so we choose to work in Fourier space, largely following the methods of \cite{Oguri:2010vi} and \cite{Jeong:2009wi}.

\subsection{Stacked Lensing Modeling}
The direct observable of stacked lensing measurements, mean tangential shear $\left<\gamma_t^h(\theta)\right>$, is directly related to the halo-shear power spectrum $C_{h\kappa}(l)$, namely
\begin{equation}
\label{eq:gammat}
\left<\gamma_t^h(\theta)\right>=\int\frac{l dl}{2\pi}C_{h\kappa}(l)J_2(l\theta).
\end{equation}
In order to work in Fourier space, we choose $C_{h \kappa}$ as our observable. \chk~is a measure of correlation between halo centers and shears. It is comprised of a small-scale (1-halo) contribution, \chkh, where halo centers and shear signals from the same halo are correlated, and a large-scale (2-halo) contribution, \chkhh, where halo centers and shear signals from different halos are correlated. The former is sourced by the density profile of halos, and the latter is sourced by the clustering of different halos. 

Galileon modifications deal with gravitational lensing arising from spherically symmetric bodies, and thus affect the 1-halo component in a well-defined manner. They specifically manifest themselves as feature-like modifications to the lensing signal at length scales dictated by halo properties, as shown in Fig. \ref{fig:plots/modfuncs}. Thus, stacking similar halos together results in stacking similar modifications together, in a detectable way. On the other hand, this effect is smaller and potentially noisier in the stacked 2-halo component, as the correlated halo center and shear signal originate from two different halos with distinct halo properties. Thus, we modify \chkh~and conservatively treat \chkhh~as unaffected by the Galileon modification.

\Wfig{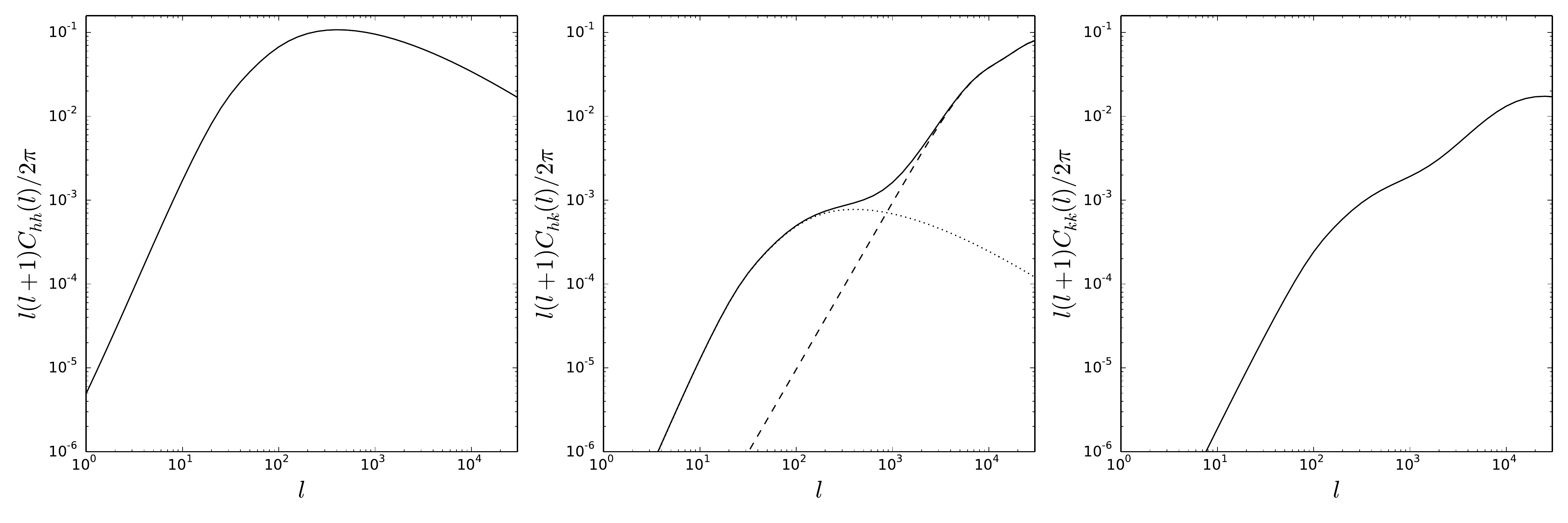}{The halo-halo, halo-shear, and shear-shear power spectra, respectively, for redshift bin $0.4<z<0.5$. For the middle panel, the 1-halo (dashed) and the 2-halo (dotted) contributions to the halo-shear power spectrum are presented.}

\subsubsection{1-halo Contribution}
The 1-halo contribution to \chk~is given by
\begin{eqnarray}
C_{h\kappa}^{1h}(l) = & & \frac{1}{n_b}\int dz\frac{d^2V}{dzd\Omega} \nonumber \\
& & \times \int dM \frac{dn}{dM}S(M,z)\tilde\kappa(l;M,c,z)
\end{eqnarray}
where $\chi$, $M$, and $c$ are the comoving distance, the halo mass, and the NFW halo concentration parameter, respectively. The spectrum depends on the lens number density $n_b$, comoving volume element per redshift per steradian $d^2V/dzd\Omega = \chi^2/H(z)$, mass function $dn/dM$, selection function $S(M,z)$, and the convergence signal $\tilde \kappa$ in Fourier space. Let us discuss each of these ingredients in turn.

The convergence signal $\tilde\kappa$ has an analytic form \cite{Scoccimarro:2000gm} derived from an NFW profile, given by
\begin{eqnarray}
\tilde \kappa(l;M,c,z) & = & \frac{M\tilde u(k=l/\chi;M,c,z)}{(1+z)^{-2}\chi^{-2}\Sigma_\text{crit}(z)},
\end{eqnarray}
where $\tilde u$ is
\begin{eqnarray}
\tilde u(k;M,c,z)  =  \frac{1}{\ln(1+c)-c/(1+c)} \left[\sin{x} \{\mathrm{Si}[x(1+c)]\frac{}{}\right. \nonumber \\
  -  \left. \mathrm{Si}{[x]}\}  +  \cos{x} \left\{\mathrm{Ci}[x(1+c)]-\mathrm{Ci}[x]\right\} - \frac{\sin (xc)}{x(1+c)}\right]. \nonumber \\
\end{eqnarray}
Here, Si and Ci are the sine and cosine integral functions, respectively, and $x\equiv(1+z)kr_s$ with $r_s$ being the NFW scale radius. The scale radius is a function of $M$ and $c$, and hence $\tilde u$ depends implicitly on $M$. We use a fitting formula for $c$ from \cite{Bullock:1999he},
\begin{equation}
c(M,z) = 7.85\left(\frac{M}{2\times 10^{12}h^{-1}M_\odot} \right)^{-0.081}(1+z)^{-0.71}.
\end{equation}
The critical surface density $\Sigma_\text{crit}$ is given by
\begin{equation}
\Sigma_{\text{crit}}^{-1}(z) = \int  d z_s p(z_s)\frac{4\pi G\chi (z)}{1+z}\left[1-\frac{\chi (z)}{\chi (z_s)}\right],
\end{equation}
with the source redshift distribution $p(z_s)$ usually modelled as
\begin{equation}
p(z_s) = \frac{z_s^2}{2z_0^3}\exp\left(-\frac{z_s}{z_0}\right),
\end{equation}
with a survey-dependent parameter $z_0$.

The selection function $S(M,z)$ defines the bins in mass and redshift. While we assume flat cuts in redshift, we follow \cite{Lima:2005tt} to consider the uncertainty in the mass selection arising from the scatter in mass-observable relations:
\begin{eqnarray}
\label{eq:selection}
S(M,z) & = &\Theta (z-z_\text{min}) \Theta (z_\text{max}-z) \nonumber \\
& \times  & \frac{[\mathrm{erfc}(y(M_\text{min}))-\mathrm{erfc}(y(M_\text{max}))]}{2}.
\end{eqnarray}
Here, $\Theta$ represents the Heaviside step function, and mass parameter $y$ is given by
\begin{equation}
\label{eq:ymass}
y(M_\text{obs}) \equiv \frac{\ln{M_\text{obs}}-\ln{M}-\ln{M_\text{bias}}}{\sqrt{2}\sigma_{\ln M}}.
\end{equation}
The redshift binning in practice should be less sharp as well due to photometric redshift errors, but that is beyond the scope of this analysis.

For the mass function $dn/dM$, we adopt the results of \cite{Bhattacharya:2010wy} and express the scaled differential mass function $f(\sigma,z)$ as 
\begin{eqnarray}
f(\sigma,z) & = & \frac{M}{\rho (z)}\frac{dn(M,z)}{d\ln{[\sigma^{-1}(M,z)]}} \nonumber \\
& = & A\sqrt{\frac{2}{\pi}}e^{-\frac{a\delta_c^2}{2\sigma^2}}\left[1+\left(\frac{\sigma^2}{a\delta_c^2}\right)^p\right]\left(\frac{\delta_c\sqrt{a}}{\sigma}\right)^q,
\end{eqnarray}
with parameters given by 
\begin{equation}
\label{eq:mfparams}
\begin{split}
A&=\frac{0.333}{(1+z)^{0.11}}, \quad a=\frac{0.788}{(1+z)^{0.01}}, \\
p&=0.807, \quad q=1.795.
\end{split}
\end{equation}
The lens number density $n_b$ is calculated by integrating this mass function weighted with the selection function. i.e.
\begin{equation}
\label{eq:nb}
n_b = \int dz \frac{d^2V}{dzd\Omega}\int dM \frac{dn}{dM}S(M,z).
\end{equation}

\subsubsection{2-halo Contribution}
The 2-halo contribution is given by
\begin{equation}
C_{h\kappa}^{2h}(l) = \int d\chi W_h(z)W_\kappa(z)\chi^{-2}P_m(k=l/\chi,z).
\end{equation}
Here, $P_m$ represents the linear matter power spectrum, with $W_h$ and $W_\kappa$ being the halo and lensing window functions, respectively. The use of the linear matter power spectrum is justified as we consider only large, and therefore linear, length scales.

The halo window function $W_h(z)$ is defined as
\begin{equation}
W_h(z) = \frac{1}{n_b}\frac{d^2V}{d\chi d\Omega}\int dM \frac{dn}{dM}S(M,z)b_h(M,z).
\end{equation}
The only new ingredient here is the halo bias $b_h$, and we again follow \cite{Bhattacharya:2010wy}  to model it as 
\begin{equation}
b_h(M,z) = 1 + \frac{a \nu - q}{\delta_c} + \frac{2p/\delta_c}{1+(a\nu)^p},
\end{equation}
where $\nu = \delta_c^2/\sigma^2$. We use the standard value of $\delta_c =1.686$, and all other parameters follow Eq. \ref{eq:mfparams}.

The lensing window function $W_\kappa(z)$ is defined as 
\begin{equation}
W_\kappa(z) \equiv \frac{\bar\rho_m(z)}{(1+z)\Sigma_{\text{crit}}(z)},
\end{equation}
with $\bar\rho_m(z)$ being the mean mass density of the universe at redshift $z$. The result of this calculation is shown in the middle panel of Fig. \ref{fig:plots/cls}.

\subsection{Modification Modeling}
\Sfig{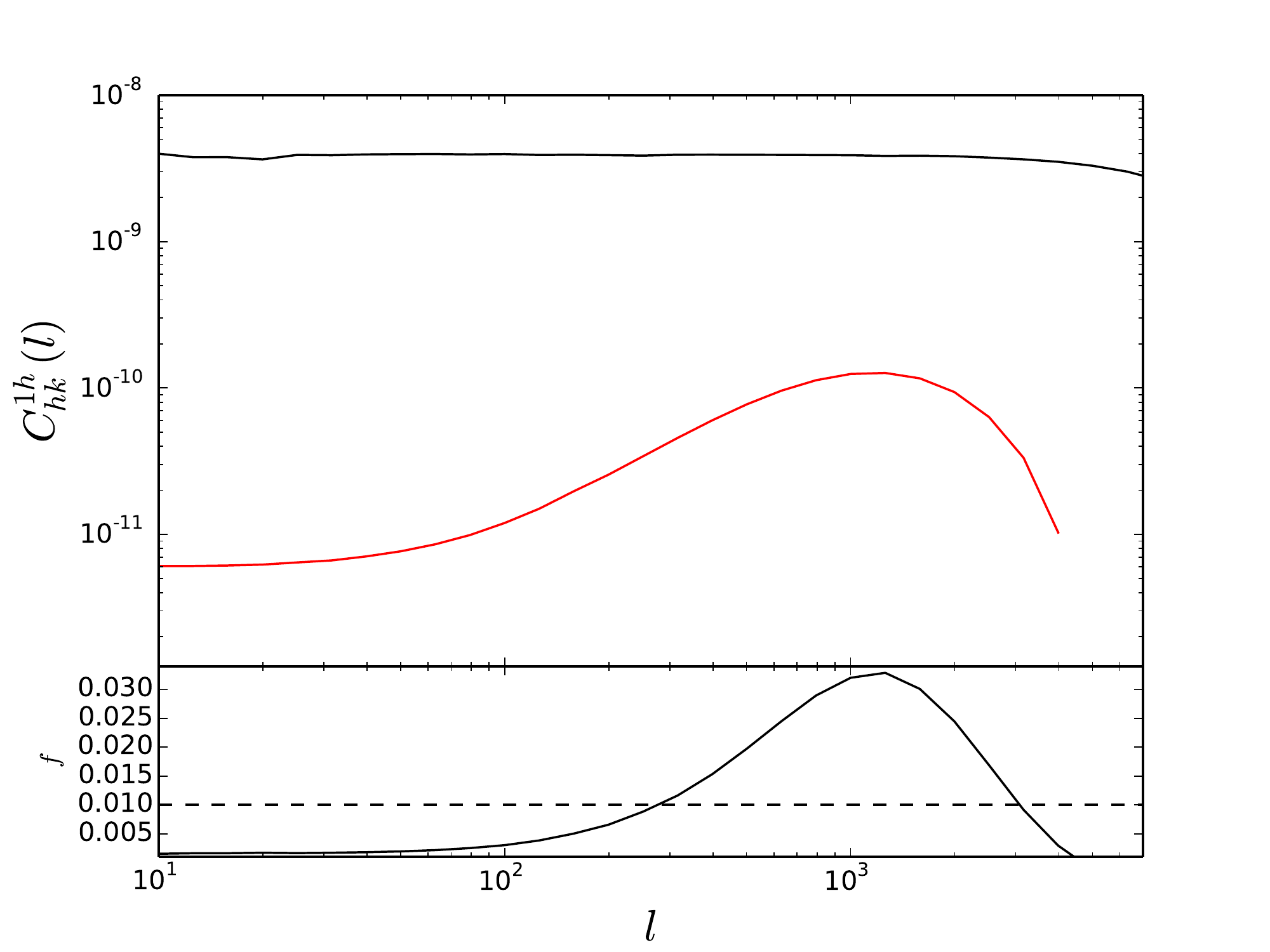}{Galileon modification for the 1-halo halo-shear power spectrum \chkh~in redshift bin $0.4<z<0.5$. Top panel shows the LCDM \chkh~(black) and its Galileon modification (red), peaking at values of $l$ corresponding to halo-sized length scales. Bottom panel shows the fractional modification to the power spectrum, with a dashed horizontal line drawn at 1\%.}

Now that the power spectra are defined, we may discuss their modification. First, the modified tangential shear signal from a halo with mass $M$ is defined as
\begin{equation}
\left<{\gamma}_t^h\right>_\text{MG} (\theta;M) = \left<\gamma_t^h\right> (\theta;M)\left(1+ \tilde{R}(r=\theta\chi;M)\right),
\end{equation}
with $\tilde{R}$ being the previously defined scaled modification function. This is a real space calculation, as $\tilde{R}$ is given in real space, and we use Eq. \ref{eq:gammat} to produce $\left<\gamma_t^h\right> (\theta;M)$. Also, note that the above expression implicitly depends on $c$. We have previously parametrized $c$ in terms of $M$ and $z$, but since the redshift dependence of $c$ within a single redshift bin is small we simply fix $z$ at each redshift bin and derive $c$ solely from $M$. This approximation is further justified given the small $c$-dependence of the modification function, as shown in Figure \ref{fig:plots/modfuncs}.

From the modified tangential shear, we obtain the modified halo-shear power spectrum for a given mass:
\begin{equation}
C_{h\kappa,\text{MG}}(l;M) = \int 2\pi l dl J_2(l\theta) \left<{\gamma}_t^h\right>_\text{MG}  (\theta;M).
\end{equation}
This is then integrated over the range of a mass bin to yield the final modified halo-shear power spectrum:
\begin{equation}
C_{h\kappa,\text{MG}}(l) = \int dM \frac{dn}{dM} S(M) C_{h\kappa,\text{MG}}(l;M),
\end{equation}
where the selection function $S(M)$ is the mass selection term from the full selection function in Eq. \ref{eq:selection}.

Figure \ref{fig:plots/frac_sig} illustrates the Galileon modification in Fourier space. The modification reaches up to 3\%, and exceeds the percent level for a relatively wide range of multipoles. The exact location and width of the peak depends on object selection, as it corresponds to a peak in real space illustrated in Figure \ref{fig:plots/modfuncs}.

\subsection{Covariance}

We adopt the Gaussian covariance for \chk~from \cite{Jeong:2009wi}, given by
\begin{equation}
\label{eq:cov}
\begin{split}
\text{Cov}&[C_{h\kappa,i}(l),C_{h\kappa,j}(l')] =  \frac{4\pi}{\Omega _s}\frac{\delta_{ll'}^K}{(2l+1)\Delta l} \\
& \times\left[\left(C_{hh,i}+\frac{1}{n_b}\right)\left(C_{\kappa\kappa,i}+\frac{\sigma_\gamma^2}{n_S}\right)\delta_{ij}^K+C_{h\kappa,i}C_{h\kappa,j}\right].
\end{split}
\end{equation}
Here, subscripts $i, j$ represent indices for redshift bins and $\Omega_s$ stands for the survey area. The quantities $n_b$, $n_S$, and $\sigma_\gamma^2$ represent the lens and source number densities and the shape noise term, respectively. We also introduce two new power spectra, \chh~and \ckk, standing for the halo-halo and the shear-shear power spectra, respectively. Let us briefly discuss the dominant components of this covariance. Sample variance, especially the \chh\ckk~ term, dominate the covariance at low multipoles. For $l>1000$, however, the cosmic shear contribution $C_{\kappa\kappa}/n_b$ dominates, and this serves as the main source of uncertainty for the Galileon signal. Because of the relatively high number densities of sources and lenses, shot noise is negligible all the way out to $l\sim 10000$. This general pattern is similar to results presented in Fig. 8 of \cite{Oguri:2010vi}, except for the shot noise contribution.

\subsubsection{Power Spectra}

The halo-halo power spectrum is given by
\begin{equation}
\label{eq:chh}
C_{hh}(l) = \int d\chi W_h(z)W_h(z)\chi^{-2}P_m(k=l/\chi,z),
\end{equation}
while the shear-shear power spectrum is given by
\begin{equation}
\label{eq:ckk}
C_{\kappa\kappa}(l) = \int d\chi W_\kappa(z)W_\kappa(z)\chi^{-2}P_m^\text{NL}(k=l/\chi,z).
\end{equation}
Eq. \ref{eq:ckk} is similar to Eq. \ref{eq:chh} in structure, modulo the switch from linear matter power spectrum $P_m$ to the non-linear matter power spectrum $P_m^{\text{NL}}$. We implement this switch by adopting HALOFIT \cite{Smith:2002dz} to extrapolate for the non-linear result from the linear result. The resulting halo-halo and shear-shear power spectra are presented in Fig. \ref{fig:plots/cls}.

\Wfig{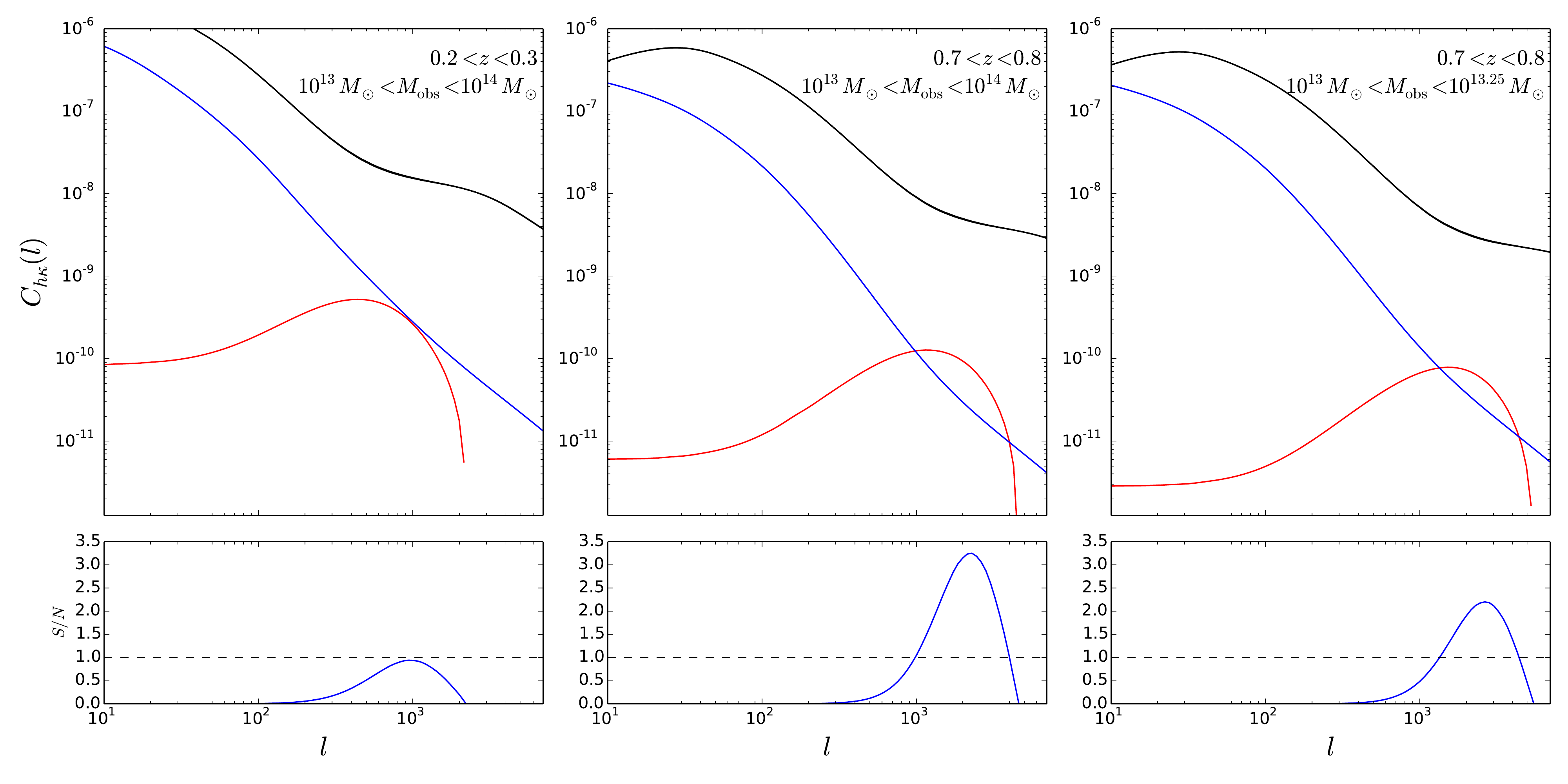}{Background (black), signal (red), and magnitude of 1-$\sigma$ errors (blue) for different redshift and mass bins. Errors are calculated assuming 10 multipole bins per decade, i.e. $\Delta \log l = 0.1$. Bottom panel shows the resulting signal-to-noise ratio. We observe improvements in signal-to-noise ratio with higher redshifts and finer mass binnings.}

\subsubsection{Survey Parameters}

In addition to the power spectra, we need to set a number of survey parameters, namely $n_b$, $n_S$, $f_\text{sky}\equiv 4\pi/\Omega_s$, and $\sigma_\gamma$ , to complete our covariance modeling. The lens number density $n_b$ is calculated as according to Eq. \ref{eq:nb}. The source number density $n_S$ can range from 10 per square arcminute in DES-like surveys \cite{Abbott:2005bi} to 50 per square arcminute in LSST-like surveys \cite{Abell:2009aa}. Sky coverage $f_\text{sky}$ is $0.12$ and $0.48$ for DES and LSST-like surveys, respectively. Finally, $\sigma_\gamma$ accounts for the shape noise, and we set it to be 0.35.

\subsection{Systematics}

In addition to the modifications of interest, there are a number of systematic effects that may alter the observed stacked lensing signal. In order to study if a Galileon modification can be distinguished from such effects, we model some of the most common systematics as follows.

\textbf{Off-centering.} In stacked lensing analyses, a common systematic effect known as off-centering drowns the convergence signal $\tilde\kappa$. As evident from the naming of the systematic, it originates from misidentification of halo centers, where one would assume a point other than the true halo center as the reference point for lensing measurements. Following \cite{Oguri:2010vi}, the off-centered convergence signal can be modeled as
\begin{equation}
\tilde\kappa_\text{off}(l) = \tilde\kappa(l)\left[f_\text{cen}+(1-f_\text{cen})\exp\left(-\frac{1}{2}\sigma_s^2l^2\right)\right],
\end{equation}
assuming a Gaussian distribution of the center of different stacked objects being dislocated with respect to one another. $\sigma_s$ is empirically determined \cite{Johnston:2007uc} to be
\begin{eqnarray}
\sigma_s & = & 0.42h^{-1}\mathrm{Mpc}/D_A(z),
\end{eqnarray}
and $f_\text{cen}$ is the fraction of halo centers correctly identified.

\textbf{Mass bias.} In Eq. \ref{eq:selection}, the parameter $y(M_\text{obs})$ involves a term $M_\text{bias}$ that accounts for a systematic bias between the observed mass and the true mass of objects. It is usually set to zero, meaning no bias in mass measurements, but we can model a simple additive mass bias by assigning a non-zero value to it. In doing so, we assume all mass values in Eq. \ref{eq:ymass} to be in units of solar mass.

\section{Detectability of Galileon Modifications}
\label{sec:results}

\subsection{Mock Survey}

In producing the power spectra, we set up four redshift bins, with $z_\text{min}=\{0.2,0.4,0.7,0.9\}$ and $\Delta z = 0.1$. We focus on halos with observed masses in the range $10^{13}M_\odot < M < 10^{14}M_\odot$. Note that the peak location of modification for an object scales with its virial radius, implying that wide mass bins will blend signals peaking at different scales to form a smoother feature that is harder to detect or differentiate from other effects. Therefore, different binning schemes within the given mass range are studied, and we bin this range with 1, 2, 4, and 10 logarithmic mass bins. LSST-like survey parameters are adopted, with $n_S$ at 50 per square arcminute and a sky coverage of 20,000 square degrees.

\subsection{Galileon Signal}

Figure \ref{fig:plots/signal} displays LCDM and Galileon \chk~for different mass and redshift bins. The Galileon ``signal", defined as the difference between the Galileon and LCDM \chk, is plotted in red. It peaks around $l\sim1000$ to $l\sim3000$, corresponding to the positive peak in our modification function. There is another negative peak at $l\geq 8000$, not shown in the plot, sourced by the small-scale dip in the modification function. However, this borders closely with the high-$l$ regime dominated by resolution issues and non-linearities, so we choose to discard this regime and focus on the positive peak. Note that the peak location coincides closely with the transition from a 2-halo dominated regime to a 1-halo dominated regime. Under certain configurations, 2-halo background and its contribution to the covariance will end up shadowing our signal altogether, as is the case for the leftmost panel. 

In order to be outside of this 2-halo ``shadow,'' the signal would have to peak at higher values of $l$, corresponding to smaller angular scales. Recall that the physical location of the signal in real space scales with the virial radius of the lens, and will stay constant with respect to redshift. Therefore, smaller masses and higher redshifts will drive down the angular scale of the peak, and signals from such objects will exhibit good separation from the 2-halo dominated regime, resulting in good $S/N$ output. Such configurations, however, are limited by survey parameters. Surveys are luminosity-limited at high redshifts, yielding increased shot noise and possibly incomplete observations. Small angular scales, e.g. $\theta < 1'$ or $l> 10000$, are easily contaminated by the blending of lensing signals with light from galactic center, as well as by limitations in survey resolution. Thus, it is unrealistic to include redshifts too high or objects too light, and we choose our redshift and mass ranges conservatively, as previously discussed.

Even with this conservative choice, we obtain optimistic results. At $0.2<z<0.3$, with the entire mass range unbinned, the signal fails to achieve $S/N>1$ anywhere, as shown in the leftmost panel. However, if we look at the center panel, at $0.7<z<0.8$, still unbinned in mass, the signal achieves $S/N>1$ for $1000 < l < 4000$, peaking at $S/N=3$. Considering the 10-per-decade binning in $l$, there will be at least two bands with $S/N>2$. Now, if we consider binning into 4 mass bins for the same redshift bin, as shown in the rightmost panel, we note that $S/N>1$ is achieved even with increased shot noise, owing to sharpening of features. This panel specifically plots the lowest of the 4 mass bins, and we observe that the peak location for these lighter objects, combined with high enough redshift, allows a safe separation of the signal from the shadow. These configurations with realtively high S/N serve as major contributions for parameter constraints in the Fisher analysis. In addition, information from cross-correlating different redshift bins, which is not affected by the dominant first term in Eq. \ref{eq:cov}, acts as a significant source of constraining power.

\Wfig{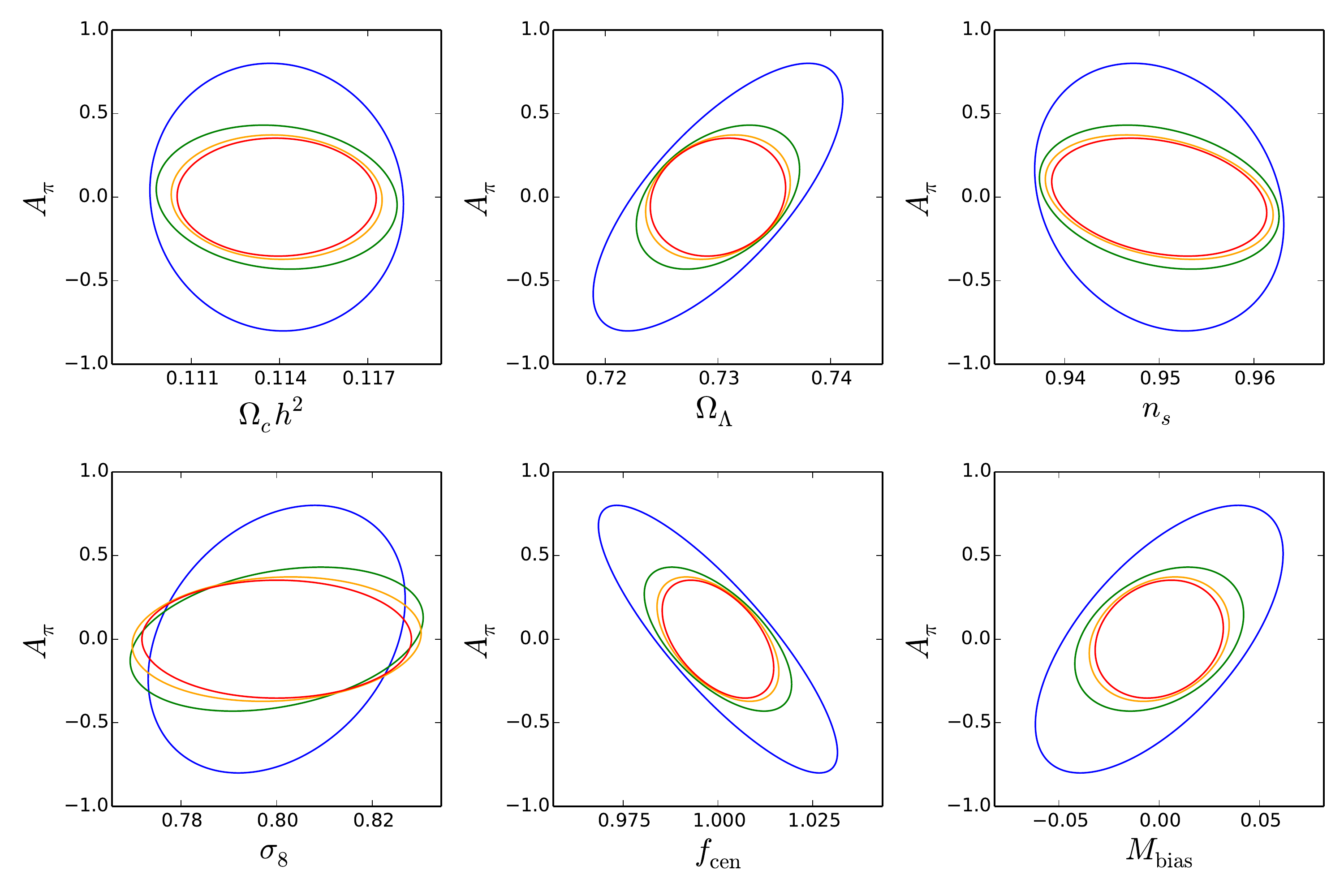}{1-$\sigma$ confidence ellipses for cosmological and nuisance parameters versus scaling parameter $A_\pi$, for 1 (blue), 2 (green), 4 (orange), and 10 (red) mass bins. With increasing number of mass bins, we observe tighter constraints on $A_\pi$, as well as less degeneracies between $A_\pi$ and other parameters.}

\subsection{Fisher Analysis}

With our observables calculated, we perform a Fisher analysis for parameter forecasts. We define our data vector as the set of $C_{h\kappa}(l)$ for the previously discussed redshift bins, i.e.
\begin{equation}
\mathbf{D} \equiv \left\{C_{h\kappa}(l)\right\}_{i,b},
\end{equation}
where $i, b$ runs over the redshift and mass bins, respectively. Then, the Fisher matrix element $\mathbf{F}_{\alpha\beta}$ corresponding to the parameters $p_\alpha$ and $p_\beta$ is given by
\begin{equation}
\label{eq:fisher}
\mathbf{F}_{\alpha\beta} = \sum_{l_\text{min}}^{l_\text{max}}\sum_{b} \frac{\partial\mathbf{D}_{b}(l)}{\partial p_\alpha} [\mathrm{Cov}(\mathbf{D}_{b}(l),\mathbf{D}_{b}(l))]^{-1} \frac{\partial\mathbf{D}_{b}(l)}{\partial p_\beta}.
\end{equation}
Here, $\mathbf{D}_b$ consists of four \chk, one for each redshift bin, for the $b$th mass bin. We consider different mass bins to be uncorrelated to each other, allowing the sum over $b$.

This setup implies that we consider LCDM as our null hypothesis, treating the Galileon modification as deviation. In order to simulate the maximum use of information on LCDM parameters obtained by surveys, we introduce priors on cosmological parameters. In particular, we adopt the error bars on cosmological parameters from Planck \cite{Ade:2013zuv}, and add it to our analysis in the form of an additional Fisher matrix.

\begin{table}[ht]
\begin{center}
\begin{ruledtabular}
\begin{tabular}{c c c c c c}
Cosmology & $\Omega_b h^2$ & $\Omega_c h^2$ & $\Omega_\Lambda$ & $n_s$ & $\sigma_8$ \\
Nuisance & $f_\text{cen}$ & $M_\text{bias}$\\
Galileon & $A_\pi$ 
\end{tabular}
\end{ruledtabular}
\caption{List of cosmological, nuisance, and Galileon parameters included in Fisher analysis.}
\label{tab:params}
\end{center}
\end{table}

Table \ref{tab:params} presents the different parameters included in the Fisher analysis. We inlcude five cosmological parameters, two nuisance parameters for off-centering and mass bias, and the scaling parameter $A_\pi$. Fiducial values for the parameter set are given by $\{\Omega_b h^2,\Omega_c h^2,\Omega_\Lambda,n_s,\sigma_8,f_\text{cen}, M_\text{bias}, A_\pi\} = \{0.02218, 0.1139, 0.73, 0.96, 0.8, 1, 0, 0\}$. This implies that we assume a perfectly centered stacking with zero mass bias as our default scenario. For density parameters, flatness of the universe is preserved by first varying a parameter presented above and then adjusting $h$ such that $\Omega_b + \Omega_c + \Omega_\Lambda = 1$. The rest of the parameters are varied trivially. It is also worth noting that in Eq. \ref{eq:fisher} the lower and upper bounds for summation in multipole must be defined. We set $l_\text{max}=8000$, as multipoles higher than this bring little change to the Fisher analysis while easily dominated by non-linearities, baryonic physics, or low resolution. We also set $l_\text{min}=8$, a safe value considering sky coverage of future experiments such as DES or LSST. Fisher analysis results are relatively insensitive to $l_\text{min}$, as there is little Galileon signal to be found at such a low multipole regime.

Figure \ref{fig:plots/ellipses} presents the results of the Fisher matrix analysis. The behavior of $\Omega_b h^2$ is almost identical to that of $\Omega_c h^2$, and was therefore omitted from the plots. In Table \ref{tab:results}, we present the marginalized standard error on $A_\pi$ as $\sigma (A_\pi)$, as well as the Pearson correlation coefficients between $A_\pi$ and other parameters as $\rho (p_i)$, for different numbers of mass bins ($N_\text{bin}$). $\sigma(A_\pi)$ shows significant improvements with increasing $N_\text{bin}$, and it is noteworthy that going from one to two mass bins allows a two-fold increase in constraining power. Furthermore, some significant parameter degeneracies observed with $N_\text{bin} = 1$, notably with the density parameters and the miscentering parameter $f_\text{cen}$, are well broken by $N_\text{bin}=4$.

\begin{table}[tbp]
\begin{center}
\begin{ruledtabular}
\begin{tabular}{c  c c c c c c c}
$N_\text{bin}$ & $\sigma (A_\pi)$ & $\rho (\Omega _c)$ & $\rho (\Omega_\Lambda)$ & $\rho (n_S)$ & $\rho (\sigma _8)$ &$\rho (f_\text{cen})$ & $\rho(M_\text{bias})$\\
1 & 0.527 & -0.501 & 0.724 & -0.207 & 0.297 & -0.847 & 0.638  \\ 
2 & 0.283 & -0.109 & 0.387 & -0.266 & 0.300 & -0.626 & 0.342 \\
4 & 0.245 & -0.044 & 0.232 & -0.286 & 0.097 & -0.522 & 0.227 \\
10 & 0.232 & -0.015 & 0.160 & -0.264 & 0.001 & -0.488 & 0.189
\end{tabular}
\end{ruledtabular}
\caption{Standard errors on $A_\pi$ and Pearson correlation coefficients for cosmological and nuisance parameters for different numbers of mass bins ($N_\text{bin}$). }
\label{tab:results}
\end{center}
\end{table}

\section{Discussion}
\label{sec:discussion}

Future cosmological surveys offer bright propspects for expanding our understanding of cosmology, including a unique window for testing theories of modified gravity. We have shown that precise measurements of weak lensing provided by such surveys may be used to distinguish Galileon theories from GR, specifically by looking for modifications in stacked measurements of galaxy-galaxy lensing. Our results suggest that redshift tomography and finer mass binning significantly increase our ability to constrain the value of the scaling parameter $A_\pi$, thereby allowing us to tell between GR and Galileon theories with greater confidence. 

Based on our parameter forecasts, we conclude that the outlook for analyzing Galileon modifications is optimistic. This is because while our analysis significantly depends on mass binning, we have good reason to believe sufficiently fine mass bins will be possible in future surveys. Ten mass bins in a decade may be unrealistic in the near future, but four is definitely possible. For example, data from the Sloan Digital Sky Survey III has been successfully analyzed with three mass bins within our mass range of $10^{13}M_\odot < M < 10^{14}M_\odot$ \cite{More:2014uva} for cosmological parameters, and future surveys with wider sky coverage and greater galaxy number densities will certainly be capable of even finer binnings in mass. 

With four mass bins, we obtain $\sigma(A_\pi)=0.245$. Let us remark on what this number implies. As we assume the LCDM case, i.e. $A_\pi = 0$, as our null hypothesis, a clear detection of the Galileon signal, i.e. $A_\pi = 1$, can be considered as a 4-$\sigma$ detection. Turning this around, a clear non-detection, i.e. $A_\pi = 0$, will imply a similarly strong exclusion of the Galileon model, assuming that covariances largely stay the same with respect to $A_\pi$. Matters become more complicated in case of $0<A_\pi<1$, as the unscaled modification function $R(\theta)$ presented in Eq. \ref{eq:modfunc} is a firm prediction of the theory, implying that $A_\pi=1$ is required by the theory. Such a case will then allow for multiple possible explanations, such as exotic systematics or suppressed Galileon signals. However, we observe that the Galileon signal is distinct from possible features sourced by cosmological and nuisance parameters, as presented above, and thus a signal following the characteristic features of a Galileon modification is a strong hint at the existence of Galileonic effects. In conclusion, our type of analysis of an LSST-like dataset has the ability to make a definitive test of the Galileon model.

Tests of modified gravity are increasingly becoming standard and recommended \cite{Jain:2013wgs} as a part of the cosmological analysis of future surveys. DES is already taking data relevant to such tests, and will be succeeded by LSST, providing a continuum of improved measurements ripe for analysis. It is likely that we will be able to start looking for signs of modified gravity, or lack thereof, in the near future, and the results of this work will serve as a useful template to be included in the suite of tests of gravity.

\begin{acknowledgments}
We deeply thank Wayne Hu for initiating this work, and Scott Dodelson for motivating and supporting a renewed analysis. We also thank Matthew Becker, Eduardo Rozo, and Masahiro Takada for helpful discussions and comments. This work was supported in part by the Kavli Institute for Cosmological Physics at the University of Chicago through grant NSF PHY-1125897, and an endowment from the Kavli Foundation and its founder Fred Kavli. MW was supported by the James Arthur Postdoctoral Fellowship at NYU during part of the time in which this work was being completed.
\end{acknowledgments}

\bibliography{references}

\end{document}